\documentclass[12pt]{iopart}

\usepackage{graphicx}
\newcommand{\acrofig}{FIG.}

\begin{document}

\title[Novel interpretation of DW dynamics along FiM strips]{Novel interpretation of recent experiments on the dynamics of domain walls along ferrimagnetic strips}

\author{Eduardo Mart\'{\i}nez$^{1}$, V\'{\i}ctor Raposo$^{1}$ and \'{O}scar Alejos$^{2}$}
\address{$^{1}$Dpto. F\'{\i}sica Aplicada, Universidad de Salamanca, 37008 Salamanca, Spain \\ $^{2}$Dpto. Electricidad y Electr\'{o}nica, Universidad de Valladolid, 47011 Valladolid, Spain}
\ead{oscar.alejos@uva.es}
\vspace{10pt}
\begin{indented}
\item[]May 2020
\end{indented}

\begin{abstract}
Domain wall motion along ferrimagnets is evaluated using micromagnetic simulations and a collective-coordinates model, both considering two sublattices with independent parameters. Analytical expressions are derived for strips on top of either a heavy metal or a substrate with negligible interfacial Dzyaloshinskii-Moriya Interaction. The work focuses its findings in this latter case, with a field-driven domain wall motion depicting precessional dynamics which become rigid at the angular momentum compensation temperature, and a current-driven dynamics presenting more complex behavior, depending on the polarization factors for each sublattice. Importantly, our analyses provide also novel interpretation of recent evidence on current-driven domain wall motion, where walls move either along or against the current depending on temperature. Besides, our approach is able to substantiate the large non-adiabatic effective parameters found for these systems.
\end{abstract}

%
\submitto{\NT}
%
%
%

\section{Introduction}
Recent studies\cite{Kim:17,Caretta:18,Blasing:18,Siddiqui:18,Kim:19,Okuno:19} have focused the attention on ferrimagnets (FiMs) as feasible candidates to implement racetrack memories,\cite{Parkin:08} due to the high domain wall (DW) velocities, of the order of $1\frac{\mathrm{km}}{\mathrm{s}}$, that can be reached along magnetic strips.\cite{Kim:17, Caretta:18} Besides, such velocities linearly increase with the applied stimuli,\cite{Kim:17, Caretta:18, Siddiqui:18} in the form of either magnetic fields or electric currents. These features occur at a temperature near that of angular momentum compensation $T_A$, when FiMs present residual net magnetization together with a null angular momentum. FiMs, as formed by two antiferromagnetically coupled ferromagnetic (FM) sublattices (SLs), possess a magnetization dependence on temperature that can be inferred from those of each SL, $M_{s,1}\left(T\right)$ and $M_{s,2}\left(T\right)$. These temperature dependences, as plotted in the graph of \acrofig\ref{Fig:01}(a), can be described by expressions:
\begin{equation}
M_{s,i}\left(T\right)=M_{s,i}^0\left(1-\frac{T}{T_C}\right)^{a_i}\mathrm{,}
\end{equation}
where $T_C$ is the Curie temperature of the FiM, $M_{s,i}^0$ are the magnetization of each SL at zero temperature, and $a_i$ are certain exponents. Magnetizations of both FM SLs are equal at the temperature of magnetization compensation $T_M$, and this temperature can differ from $T_A$ due to distinct Land\'{e} factors $g_i$ for each SL.

Two different structures have been proposed to hold such fast dynamics. FiM strips grown on top of a heavy metal (HM), shown in \acrofig\ref{Fig:01}(b), are the first ones. The interface here provokes interfacial asymmetric exchange interactions, as the interfacial Dzyaloshinskii-Moriya interaction (iDMI), resulting in the formation of chiral DWs. Current driven domain wall motion (CDDWM) is then promoted by spin orbit torques (SOTs).\cite{Caretta:18,Blasing:18,Siddiqui:18} In the second structures, shown in \acrofig\ref{Fig:01}(c), FiM strips lie on a conventional substrate. The absence of iDMI interactions allows the formation of achiral DWs, whose dynamics have been characterized in both the field-driven and current-driven cases.\cite{Kim:17,Okuno:19} Spin-transfer torques (STTs) are thought to be responsible for the CDDWM in these systems.

Most of these experimental results have been interpreted in terms of effective models, which describe FiMs as effective FMs. Within these models, the effective gyromagnetic ratio $\gamma_i$ and Gilbert damping parameter $\alpha_i$ diverge at $T_A$.\cite{Stanciu:06} However, other experimental studies have shown that Gilbert damping remains constant over a wide temperature range around both $T_M$ and $T_A$.\cite{Kim:19} Importantly, the underlying physics governing the DW dynamics is also missing. Alternatively, a model considering FiMs as formed by two FM SLs coupled through an interlattice exchange interaction, i.e., a two sublattice model (TSLM), can be proposed.\cite{Martinez:19,Martinez:20} Two Landau-Lifshitz-Gilbert equations are posed, one for each SL, along with an exchange interaction term between the two SLs. As an advantage over effective models, fixed values for the parameters involved in the model can be chosen, based on the experimental evidence. Particularly, this work shows how the TSLM yields a novel and more plausible interpretation of the observations of very recent results\cite{Okuno:19} on CDDWM in FiMs, where this motion can occur along or against the current depending on temperature. Besides, a key to substantiate the relatively large non-adiabatic parameters found in these materials is provided by means of positive non-adiabatic parameters and different spin polarization of the two sublattices. The TSLM can be implemented using micromagnetic ($\mu$M) simulations or as a collective coordinate model (CCM) and it is briefly set out in the following section.

\section{Models}
\begin{figure}[h]
\centering
  \begin{tabular}{@{}cc@{}}
    (a)\includegraphics[scale=1]{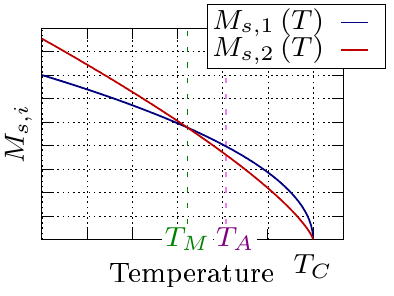}&
    (d)\includegraphics[scale=1]{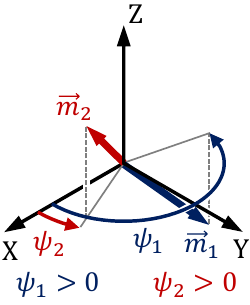} \\
    \multicolumn{2}{c}{(b)\includegraphics[scale=0.8]{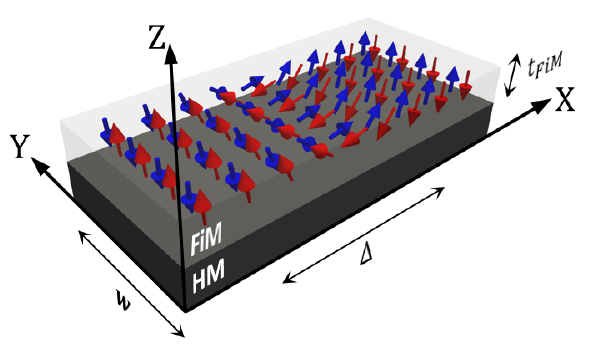}}\\
    \multicolumn{2}{c}{(c)\includegraphics[scale=0.8]{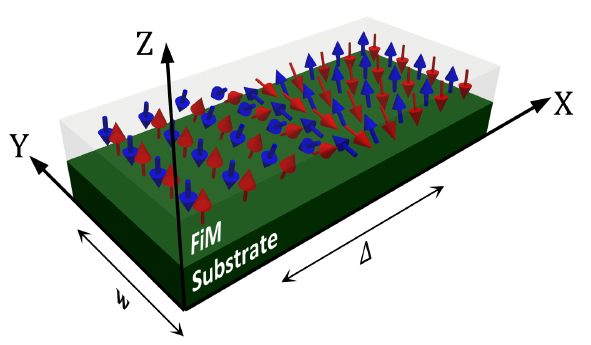}}\\
    \multicolumn{2}{c}{(e)\includegraphics[scale=0.8]{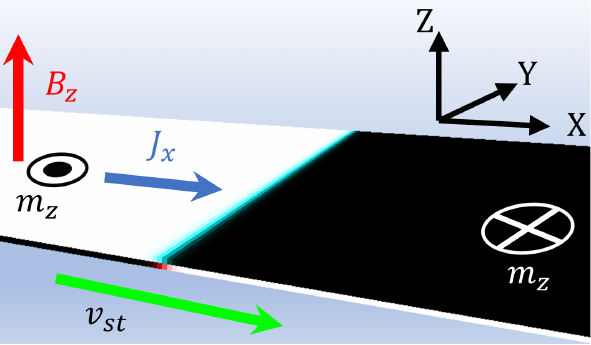}}
  \end{tabular}
\caption{Two SLs constitute the FiM: (a) temperature dependence of the magnetization of each SL, (b) magnetic DW of N{\'e}el type, and (c) magnetic DW of Bloch type amidst two domains oriented out of plane (the strip width $w$ is here shown), (d) magnetizations are represented by the unit vectors $\vec{m}_1$ and $\vec{m}_2$, with in-plane orientation angles $\psi_1$ and $\psi_2$, respectively, and (e) schematic representation of inputs and outputs of the TSLM.

Images (a), (b) and (c) reprinted from: E. Mart{\'\i}nez, V. Raposo, and {\'O}. Alejos, AIP Advances 10, 015202 (2020); licensed under a Creative Commons Attribution (CC BY) license.
}
\label{Fig:01}
\end{figure}

The system is described using two unit vectors $\vec{m}_i$, $(i=1,2)$ accounting for the local orientation of the magnetization of each SL, as shown in \acrofig\ref{Fig:01}(d). The in-plane components of these vectors provide information about the orientation of magnetic moments within the DW through the $\psi_i$ angles. The uniaxial anisotropy promotes the formation of domains with out-of-plane magnetization, and the magnetization transitions within any DW can be either from up-to-down (UD) or from down-to-up (DU). \acrofig\ref{Fig:01}(e) depicts schematically a DW in the system. Both SL are represented, one on top of the other. The transition is of UD (DU) type for the SL on top (below), due to the antiferromagnetic coupling. The individual or combined application of out-of-plane fields $B_z$ and longitudinal currents $J_x$ promote DW displacements along the longitudinal direction of the FiM strip, with velocity $v$. During the dynamics, the magnetic moments within the DW change their orientations and can reach either a steady orientation or a precessing behavior. Importantly, the complete TSLM admits misalignments between SLs, of particular interest when steady orientations are reached.\cite{Martinez:19} Nevertheless, a reduced version of the TSLM is to be adopted here, since the large antiferromagnetic coupling between SLs promotes $\psi_2+\pi\approx\psi_1=\psi$. Besides, precession of magnetic moments occurs with identical steady frequencies, i.e., $\dot\psi_{1,st}=\dot\psi_{2,st}=\omega_{st}$.

In this reduced version, $B_z$ and $J_x$ are the model inputs, and $v$ and $\omega_{st}$ are its outputs. To distinguish between the effect of STTs and SOTs, the terms $J_{FiM}$, standing for the current through the FiM, in the former case, and $J_{HM}$, representing the current through the HM, in the latter, will be used instead of $J_x$. Details of the model are in the appendices A to D and elsewhere.\cite{Martinez:19,Martinez:20} The following CCM expressions are derived:

\begin{eqnarray}
\fl  \left(\alpha_1s_1+\alpha_2s_2\right)\frac{v}{\Delta}+Q\delta_s\omega_{st}= 
    -\left(\beta_1p_1+\beta_2p_2\right)\frac{J_{FiM}}{\Delta}
    -Q\frac{\pi}{2}\frac{q_1+q_2}{t_{FiM}}J_{HM}\cos\psi+\nonumber\\
        +QMB_z\mathrm{,}\label{eqn:01}\\
\fl -Q\delta_s\frac{v}{\Delta}+\left(\alpha_1s_1+\alpha_2s_2\right)\omega_{st}=
    Q\left(p_1-p_2\right)\frac{J_{FiM}}{\Delta}+\frac{1}{2}B_mM\sin2\psi
    +QB_DM\sin\psi\mathrm{.}\label{eqn:02}
\end{eqnarray}

The parameters involved in these expressions are largely used in the literature: the net magnetization $M=M_{s,1}-M_{s,2}$, $M_{s,i}$ $(i=1,2)$ being the saturation magnetization for each SL, $\gamma_i=\frac{g_i\mu_B}{\hbar}$ are the gyromagnetic ratios, $\mu_B$ and $\hbar$ being respecitvely the Bohr magneton and the Plank constant, $s_i=\frac{M_{s,i}}{\gamma_i}$ are the SL angular momenta, $\delta_s=s_1-s_2$ is the net angular momentum, $p_i$ are the spin conversion factors,\cite{Okuno:19} related to the STT spin polarization values $P_i$ through $p_i=\frac{\hbar}{2e}P_i$, $e$ is the absolute electron charge, $\beta_i$ are the non-adiabatic STT parameters, $q_i$ are equivalent to $p_i$ for SOTs, that is, $q_i=\frac{\hbar}{2e}\theta_{SH,i}$, where $\theta_{SH,i}$ are the spin-Hall angles. $Q$ defines the DW transition type in the first SL, i.e., $Q=+1(-1)$ for an UD(DU) transition, and $t_{FiM}$ is the FiM thickness, as shown in \acrofig\ref{Fig:01}(b). $\Delta$, representing the DW width (see \acrofig\ref{Fig:01}(b) and (c)), can be taken as an additional parameter, or obtained from analytical estimations.\cite{STejerina:20} Finally, two additional fields must be defined at this point: $B_m$, representing the magnetostatic interactions, and $B_D$, which accounts for the asymmetric exchange interactions. According to the discussion in \ref{app:03}, $B_m=\mu_0\left(N_x-N_y\right)M$ and $B_D=\frac{\pi}{2}\frac{D_1+D_2}{M\Delta}$, where $N_x$ and $N_y$ are respectively the longitudinal and transverse demagnetizing factors for the DW, and $D_i$ represent the iDMI constants for each SL.

The $\mu$M simulations presented in this work have been carried out with a homemade code\cite{Alejos:18} implemented on graphic processing units (GPUs), whereas the CCM version has been implemented by means of a conventional programming language.
A comprehensive list of the parameters used for the simulations is presented below. The choice of these parameters has been made based on the values found in the literature.\cite{Caretta:18} Except if the contrary is indicated, the listed values have been adopted for all simulations. The values $T_C=450\mathrm{K}$, $M_{s,1}^0=1.4\frac{\mathrm{MA}}{\mathrm{m}}$, $M_{s,2}^0=1.71\frac{\mathrm{MA}}{\mathrm{m}}$,$a_1=0.5$, and $a_2=0.76$ have been considered. Land{\'e} factors for each SL are $g_1=2.2$ and $g_2=2$. Accordingly, the compensation temperature and the temperature of angular momentum compensation can be estimated as $T_M=241.5\mathrm{K}$ and $T_A=305\mathrm{K}$, respectively. Besides, the following values have been taken for both SLs: Gilbert damping constants are $\alpha_i=0.02$,\cite{Kim:19} exchange interactions are set to $A_i=70\frac{\mathrm{pJ}}{\mathrm{m}}$, and out-of-plane crystalline anisotropy constants take the value $k_{u,i}=1.5\frac{\mathrm{MJ}}{\mathrm{m}^3}$. According to the discussion in \ref{app:04}, the domain wall width can be estimated from these parameters as $\Delta\approx 7\mathrm{nm}$. The last common parameter for all simulations is the antiferromagnetic coupling between SLs, given by an interlattice exchange constant $B_{12}=9\frac{\mathrm{MJ}}{\mathrm{m}^3}$.\cite{Ma:16}

The following parameters are specific for CDDWM simulations.\cite{Martinez:20} The non-adiabatic STT parameters have been made equal to the Gilbert damping, i.e., $\beta_i=0.02$. The simulations where both polarization factors are equal have been carried out with $P_1=P_2=0.7$. Oppositely, the simulations where both polarization factors differ each other have been carried out with $P_1=0.8$ and $P_2=0.6$.

\section{Results}

\subsection{Ferrimagnets grown on top of a heavy metal}\label{sec:3.1}
This section has been included to give some completeness to the work, since most findings concerning systems dominated by spin-orbit torques have already been published by our group.\cite{Martinez:19} In particular, if iDMI exists, $B_D\gg B_m$ usually, and the TSLM confirms SOTs and iDMI as responsible for the CDDWM in FiM strips grown on top of a HM.\cite{Caretta:18,Blasing:18,Siddiqui:18,Martinez:19} Dynamics are characterized by the motion of rigid DWs, i.e., $\omega_{st}\equiv 0$. From (\ref{eqn:01}) and (\ref{eqn:02}), DW steady velocities $v_{st}$ are reached, which are given by the expression:
\begin{equation}
v_{st}=\frac{\pi}{2}\frac{J_{HM}}{\sqrt{\left(\frac{\alpha_1s_1+\alpha_2s_2}{q_1+q_2}\frac{t_{FiM}}{\Delta}\right)^2+\left(\frac{\pi}{2}\frac{\delta_sJ_{HM}}{\Delta B_DM}\right)^2}}\mathrm{.}
\end{equation}
The above equation, not included in our previous works, constitutes a closed-form expression that summarizes DW dynamics in these systems. According to this result, the $v_{st}$ saturate as the electric current is increased, except at $T_A$ ($\delta_s$ vanishes), where the $v_{st}$ linearly increase with $J_{HM}$ because DW magnetic moments keep aligned with the electric current. DW terminal velocities maximize at this temperature, so that $v_{st}=\frac{\pi}{2}\frac{q_1+q_2}{\alpha_1s_1+\alpha_2s_2}\frac{\Delta}{t_{FiM}}J_{HM}$.

\subsection{Ferrimagnets grown on top of a substrate with no iDMI}
\begin{figure}[ht]
\centering
\includegraphics[scale=1]{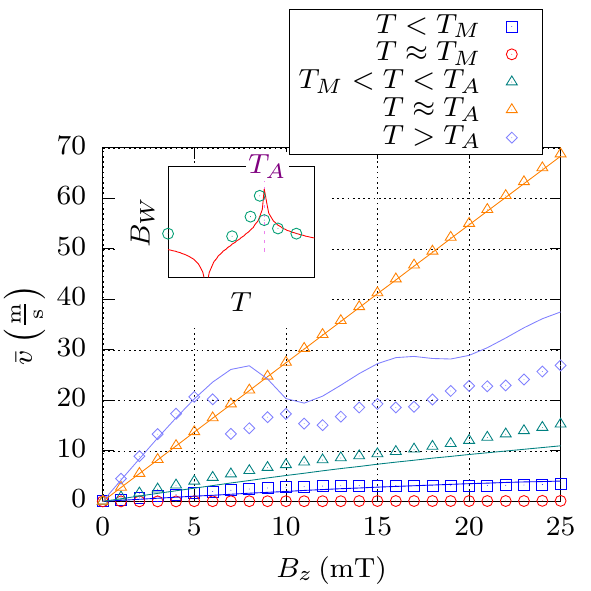}
\caption{Dependence of $\bar{v}$ as functions of $B_z$ at different temperatures. Absolute values of $\bar{v}$ have been considered since the net magnetization change sign below and above $T_M$, resulting respectively in negative and positive velocities if the sign of the net magnetization is taken into account. Dots represent the values obtained from $\mu$M simulations and continuous lines correspond to CCM results. The change in slope of the curves reveals the transition from a non-precessional to a precessional dynamic at a certain applied field $B_W$. This fact defines the condition posed to determine $B_W$ as the points at which the linearity between velocities and applied stimuli disappears. Such points are close to local velocity maxima. The inset compares the $B_W$ (in log scale) obtained from $\mu$M simulations (dots) and the CCM (continuous lines).}
\label{Fig:02}
\end{figure}

More intriguing is the case of the second FiM structure shown in \acrofig\ref{Fig:01}(c), when iDMI vanishes. Dynamics, both field-driven and current-driven, are now dominated by precessional regimes. Mean values of $v$ $\left(\bar{v}\right)$ must be considered within these regimes. Precessional regimes are due to rather low magnetostatic interactions, as discussed below. Breakdown conditions can be derived from (\ref{eqn:01}) and (\ref{eqn:02}). For applied fields, field-driven DW motion (FDDWM) results in precessional DW displacements if the field $B_z$ is above the value:
\begin{equation}
B_W=2\frac{\alpha_1s_1+\alpha_2s_2}{\left|\delta_s\right|}B_m\mathrm{.}
\end{equation}
This expression recalls the Walker field for FMs,\cite{Thiaville:05b} but adds a factor depending inversely on $\delta_s$. The expression generalizes that obtained in the literature from an effective model,\cite{Oh:17} where Gilbert damping values are taken as identical for both SL, and diverge at $T_A$. \acrofig\ref{Fig:02} presents $\bar{v}$ as functions of $B_z$ at different temperatures, obtained with the set of parameters detailed above. Only at $T_A$, $\bar{v}$ increases linearly with $B_z$. At other temperatures, the slopes of the curves reduce when the field exceeds the corresponding $B_W$ at the given temperature. Resulting $B_W$ values are of a few mT or less, except at $T_A$, where $B_W$ diverges, as schematically shown in the inset plot.

A similar analysis can be performed regarding CDDWM. In particular, the threshold current $J_W$ leading to precessional CDDWM can be worked out as:
\begin{equation}
J_W=\frac{2\left(\alpha_1s_1+\alpha_2s_2\right)M\Delta}{\left|\delta_s\left(\beta_1p_1+\beta_2p_2\right)+\left(\alpha_1s_1+\alpha_2s_2\right)\left(p_1-p_2\right)\right|}B_m\mathrm{.}\label{Eqn:06}
\end{equation}
This expression is fully consistent with that of FMs.\cite{Thiaville:05b} Indeed, the TSLM provides the correct expression for FMs if ferromagnetic coupling between SLs is considered. 

\begin{figure}[hb]
\centering
\includegraphics[scale=1]{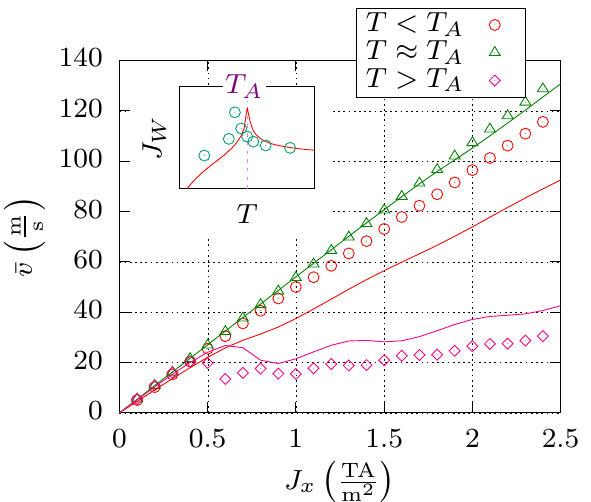}
\caption{Dependence of absolute $\bar{v}$ as functions of the longitudinal current $J_x$ at different temperatures in the case $\left(p_1=p_2\right)$. CDDWM shares in this case rather similar features to those of FDDWM. The inset presents $J_W$ (in log scale) as a function of temperature.}
\label{Fig:03}
\end{figure}

Two scenarios can be observed here. First, if both polarization values are equal $\left(P_1=P_2\right)$, the denominator of (\ref{Eqn:06}) vanishes at $T_A$, resulting in CDDWM sharing the same characteristics than FDDWM around this temperature. The CDDWM lacks of adiabatic contributions, as defined by Okuno and coworkers,\cite{Okuno:19} i.e., the term multiplying $J_{FiM}$ in (\ref{eqn:02}). This is shown in \acrofig\ref{Fig:03}. Absolute values of $\bar{v}$ are presented because CDDWM runs in the opposite direction to the electric current with the adopted set of parameters (positive non-adiabatic STTs).

\begin{figure}[ht]
\centering
\includegraphics[scale=1]{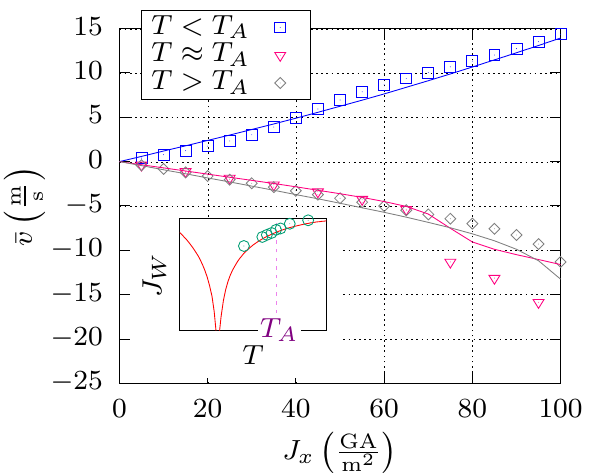}
\caption{Dependence of $\bar{v}$ as functions of the longitudinal current $J_x$ at different temperatures in the case $\left(p_1\neq p_2\right)$. Depending on temperature, CDDWM runs in the same or the opposite direction to the electric current. The inset presents $J_W$ (in log scale) as a function of temperature.}
\label{Fig:04}
\end{figure}

When $\left(P_1\neq P_2\right)$, $J_W$ does not diverge in general, even at $T_A$, due to the adiabatic STT contributions. Rather low electric currents promote DW precessional behavior. $J_W$ is of the order of one ten of $\frac{\mathrm{GA}}{\mathrm{m}^2}$ around $T_A$ for the set of parameters used. More interestingly, CDDWM can take place in either the same or the opposite direction of the electric current depending on temperature. All this is depicted in \acrofig\ref{Fig:04}.

Previous analysis demonstrates that precessional regimes dominate DW motion in FiMs grown on substrates with no iDMI in most practical situations. Within this regime $\bar{v}$ and $\omega_{st}$ can be estimated from (\ref{eqn:01}) and (\ref{eqn:02}) as:
\begin{eqnarray}
\frac{\bar{v}}{\Delta} &=& Q\frac{\left(\alpha_1 s_1+\alpha_2 s_2\right)M}{\left(\alpha_1 s_1+\alpha_2 s_2\right)^2+\delta_s^2}B_z-\label{eqn:07}\\*
&-&\frac{\left(\alpha_1 s_1+\alpha_2 s_2\right)\left(\beta_1p_1+\beta_2p_2\right)+\delta_s\left(p_1-p_2\right)}{\left(\alpha_1 s_1+\alpha_2 s_2\right)^2+\delta_s^2}\frac{J_{FiM}}{\Delta}\mathrm{,}\nonumber\\
\omega_{st} &=& \frac{\delta_sM}{\left(\alpha_1 s_1+\alpha_2 s_2\right)^2+\delta_s^2}B_z+\label{eqn:08}\\*
&+&Q\frac{\left(\alpha_1 s_1+\alpha_2 s_2\right)\left(p_1-p_2\right)-\delta_s\left(\beta_1p_1+\beta_2p_2\right)}{\left(\alpha_1 s_1+\alpha_2 s_2\right)^2+\delta_s^2}\frac{J_{FiM}}{\Delta}\mathrm{.}\nonumber
\end{eqnarray}

The expressions present linear relationships between inputs and outputs. As a first outcome, these results prove that FDDWM shows $\bar{v}$ maxima approximately when the net angular momentum vanishes, i.e., at $T_A$, as evidenced by other authors.\cite{Kim:17} Besides, these expressions also show that precession vanishes at this temperature.

The second conclusion that can be extracted from (\ref{eqn:07}) and (\ref{eqn:08}) is that the most recent results presented by Okuno and coworkers\cite{Okuno:19} can be naturally explained with the use of the TSLM. In this work, the authors distinguish adiabatic and non-adiabatic components of the CDDWM, related respectively to effective Gilbert damping $\alpha$ and non-adiabatic STT $\beta$ parameters. According to our modeling, the origin of that adiabatic term is the different spin polarizations promoted by the components of each SL.\cite{Meservey:94} The effective parameters in Okuno's model can be then obtained from these of the TSLM as: $\alpha=\frac{\alpha_1 s_1+\alpha_2 s_2}{s_1+s_2}$ and $\beta=\frac{\beta_1 p_1+\beta_2 p_2}{p_1-p_2}$. While the effective $\alpha$ is a weighted value of the $\alpha_i$ for each SL, and so, rather close to the latter values, the value $\beta$ diverges if the spin polarizations for each SL are quite close. This accounts for the large $\beta$ needed to explain CDDWM in FiMs from the perspective of effective models.

\begin{figure}[ht]
\centering
\includegraphics[scale=1]{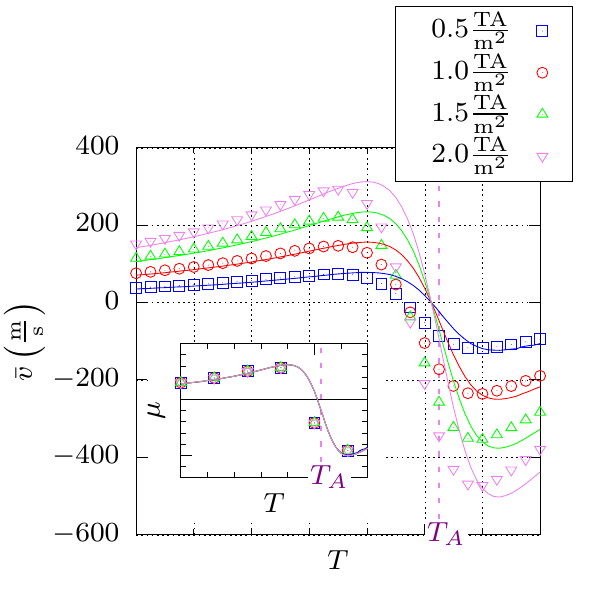}
\caption{Dependence of $\bar{v}$ as functions of temperature for different applied currents in the case $\left(p_1\neq p_2\right)$. Depending on temperature, CDDWM runs in the same or the opposite direction to the electric current. The inset presents the DW mobility $\mu$ as a function of temperature.}
\label{Fig:05}
\end{figure}

With the TSLM, general expressions for the DW mobility over the full temperature range can also be obtained. Mobilities are defined as the ratios between $\bar{v}$ and the inputs $B_z$ and $J_{FiM}$. Despite being both stimuli simultaneously applied in that experimental work, constant mobilities can be separately studied and analyzed, due to the linear character of the system, as (\ref{eqn:07}) proves. The attention will be now focused on CDDWM, and its corresponding mobility term $\mu$. This term is calculated in the experimental work as $\mu=\frac{\bar{v}\left(B_z,J_{FiM}\right)-\bar{v}\left(B_z,-J_{FiM}\right)}{2J_{FiM}}$. Because of the linear behavior, this term can be simply computed as $\mu=\frac{\bar{v}\left(0,J_{FiM}\right)}{J_{FiM}}=-\frac{\left(\alpha_1 s_1+\alpha_2 s_2\right)\left(\beta_1p_1+\beta_2p_2\right)+\delta_s\left(p_1-p_2\right)}{\left(\alpha_1 s_1+\alpha_2 s_2\right)^2+\delta_s^2}$. Importantly, this parameter changes sign depending on temperature. Accordingly, CDDWM reverses around a temperature in the vicinity of $T_A$, i.e., DW motion takes place in the same or the opposite direction to the electric current depending on temperature. This is shown in \acrofig\ref{Fig:05}, where $\bar{v}$ are plotted as functions of temperature for different electric currents. At temperatures below (above) approximately $T_A$, DW motion takes place along (oppositely) the direction of the electric current. Constant mobility can be checked in the inset plot, where $\mu$ is computed in the four cases considered, since all computed graphs accurately superpose. The curves plotted in \acrofig\ref{Fig:05} adequately emulates the aforementioned experimental results.\cite{Okuno:19} It must noted that together with different polarization factors for each sublattice, our results were obtained by assuming positive non-adiabatic parameters. Indeed, the explanation of the experimental results\cite{Okuno:19} were based on an effective model that requires a negative non-adiabatic parameter, which lacks of experimental verification.

\section{Conclusions}
To conclude, the present analysis highlights the capabilities of the TSLM to explain very recent evidence on DW motion in FiM strips. Two different structures have been considered. In the case of FiMs grown on top of a HM, where SOTs dominate the CDDWM, a brief review of the relevant features of these dynamics is made. In these first structures, $v_{st}$ present maxima when net angular momenta vanishes at $T_A$. The second structure, where FiMs are grown on a substrate with no iDMI, is characterized by richer DW dynamics. Precession dominates such dynamics in most practical situations. In particular, the angular momentum compensation results in $\bar{v}$ maximizing around $T_A$ for the FDDWM. Besides, CDDWM presents adiabatic and non-adiabatic contributions. The adiabatic term results from the different spin polarizations\cite{Meservey:94} promoted by the components of either SL. As this difference reduces, the adiabatic term vanishes. This fact explains why large non-adiabatic parameters are required to interpret the experimental results in these structures from the perspective of effective models. The prevalence of the non-adiabatic term over the adiabatic one results in CDDWM more similar to FDDWM. Last but not least, the TSLM has succesfully interpreted novel experiments, where CDDWM can take place in one or the opposite direction depending on temperature. This change in the dynamics occurs near $T_A$. Accordingly, the TSLM has been proved to be a primary tool to deduce in the near future the key material parameters governing experimental observations.

\ack
This work was partially supported by Project No. MAT2017-87072-C4-1-P from the (Ministerio de Econom{\'\i}a y Competitividad) Spanish Government and Project No. SA299P18 from the (Consejer{\'\i}a de Educaci{\'o}n of) Junta de Castilla y Le{\'o}n.

The data that supports the findings of this study are available within the article.

\appendix

\section{Calculation of the demagnetizing energy}\label{app:01}
The demagnetizing energy in the TSLM is calculated from the description of the magnetization textures by means of some collective coordinates. Such collective coordinates are the instantaneous position $q$ of a domain wall (DW) in the system, and the orientations $\psi_i$ of the in-plane components of the DW moments of each sublattice (SL) with respect to the longitudinal axis (X-axis) of the strip. The local magnetization of each SL, given by their respective saturation values $M_s$ and two unit vectors $\vec{m}_i$ defining their local orientation by means of their polar angles $\theta_i$ and azimuthal angles $\phi_i$, is then written as functions of these collective coordinates through the well-known ansatz $\theta_i=2Q_i\arctan\left(\frac{x-q}{\Delta}\right)$ and $\phi_i=\psi_i$, $\Delta$ accounting for the DW width, and $Q_i$ determining the magnetization transition, as they have been defined in the main text. The approach that both SLs share the same position $q$ and width $\Delta$ has been made here, which seems to be valid even for slightly weak coupling between SLs.\cite{Alejos:18} Additionally, $\Delta$ can be estimated from the other model parameters, as it will be discussed in \ref{app:04}.
Before the application of the ansatz, the demagnetizing energy can be expressed as:
\begin{equation}
\fl
\epsilon_m=\frac{1}{2}\mu_0\left(M_{s,1}\vec{m}_1+M_{s,2}\vec{m}_2\right)\left(
\begin{array}{c c c}
N_x & 0 & 0\\
0 & N_y & 0\\
0 & 0 & N_z
\end{array}
\right)
\left(M_{s,1}\vec{m}_1+M_{s,2}\vec{m}_2\right)^T\mathrm{,}\label{eqn:A1}
\end{equation}
$\mu_0$ being the vacuum permeability and $N_x$, $N_y$ and $N_z$ representing the demagnetizing factors given by the DW dimensions. From (\ref{eqn:A1}) and the use of the ansatz, the density energy per unit area $\sigma_m$ is computed as:
\begin{eqnarray}
\fl
\sigma_m=\int_{-\infty}^{\infty}\epsilon_mdx=\mu_0\Delta\left[N_x\left(M_{s,1}\cos\psi_1+M_{s,2}\cos\psi_2\right)^2+\right.\nonumber\\
\left.+N_y\left(M_{s,1}\sin\psi_1+M_{s,2}\sin\psi_2\right)^2-N_z\left(M_{s,1}+Q_1Q_2M_{s,2}\right)^2\right]\mathrm{,}
\end{eqnarray}
where the results $\int_{-\infty}^{\infty}\sin^2\theta_i dx=2\Delta$ and $\int_{-\infty}^{\infty}\cos^2\theta_i dx=-2\Delta$ have been considered. For antiferromagnetic (ferromagnetic) coupling, it follows that $Q_1Q_2 = -1\left(+1\right)$.

\section{Calculation of the iDMI energy.}
The energy density per unit area accounting for the asymmetric exchange interactions, as it is the case of the interfacial Dzyaloshinskii-Moriya interaction (iDMI), is derived in a similar fashion to the demagnetizing energy. The calculation starts from the energy density:
\begin{equation}
\fl\epsilon_D=D_1\left[\left(\hat{z}\cdot\vec{m}_1\right)\nabla\vec{m}_1-\left(\vec{m_1}\nabla\right)\left(\hat{z}\cdot\vec{m}_1\right)\right]+D_2\left[\left(\hat{z}\cdot\vec{m}_2\right)\nabla\vec{m}_2-\left(\vec{m_2}\nabla\right)\left(\hat{z}\cdot\vec{m}_2\right)\right]\mathrm{,}\label{eqn:B1}
\end{equation}
where $D_i$ represent the iDMI constants for each SL, and $\hat{z}$ is the unit vector in the out-of-plane direction. Integration along the longitudinal axis of (\ref{eqn:B1}) and the use of the ansatz results in the following area density:
\begin{equation}
\sigma_D=\pi Q_1D_1\cos\psi_1+\pi Q_2D_2\cos\psi_2\mathrm{.}
\end{equation}

\section{Derivation of the simplified collective coordinate model (CCM) for the TSLM}\label{app:03}
The minimization of the whole functional including $\sigma_m$ and $\sigma_D$ goes through the calculation of its derivatives with respect to $q$, $\dot{q}$, and $\psi_i$ and $\dot\psi_i$ (more details can be found in the literature\cite{Martinez:19}). This results in a system of the following three equations that make up the generalized CCM for the TSLM.
\begin{eqnarray}
\fl \frac{\dot q}{\Delta}\left(\alpha_1s_1+\alpha_2s_2\right)+Q_1s_1\dot\psi_1+Q_2s_2\dot\psi_2=\nonumber\\
    =-\left(\beta_1p_1+\beta_2p_2\right)\frac{J_{FiM}}{\Delta}+\left(Q_1M_{s,1}+Q_2M_{s,2}\right)B_z-\nonumber\\
    -\frac{\pi}{2}\left(Q_1q_1\cos\psi_1+Q_2q_2\cos\psi_2\right)\frac{J_{HM}}{t_{FiM}}\mathrm{,}\label{eqn:C1}\\
\fl -Q_1\frac{\dot{q}}{\Delta}s_1+\alpha_1\dot\psi_1s_1=\nonumber\\
    =Q_1p_1\frac{J_{FiM}}{\Delta}-\frac{1}{2}\mu_0\left(N_y-N_x\right)M_{s,1}^2\sin 2\psi_1-\nonumber\\
    -\mu_0M_{s,1}M_{s,2}\left(N_y\cos\psi_1\sin\psi_2-N_x\sin\psi_1\cos\psi_2\right)+\nonumber\\
    +Q_1\frac{\pi}{2}\frac{D_1}{\Delta}\sin\psi_1-B_{ex}\sin\left(\psi_1-\psi_2\right)\mathrm{,}\label{eqn:C2}\\
\fl -Q_2\frac{\dot{q}}{\Delta}s_2+\alpha_2\dot\psi_2s_2=\nonumber\\
    =Q_2p_2\frac{J_{FiM}}{\Delta}-\frac{1}{2}\mu_0\left(N_y-N_x\right)M_{s,2}^2\sin 2\psi_2-\nonumber\\
    -\mu_0M_{s,1}M_{s,2}\left(N_y\sin\psi_1\cos\psi_2-N_x\cos\psi_1\sin\psi_2\right)+\nonumber\\
    +Q_2\frac{\pi}{2}\frac{D_2}{\Delta}\sin\psi_2+B_{ex}\sin\left(\psi_1-\psi_2\right)\mathrm{,}\label{eqn:C3}
\end{eqnarray}
where all parameters have already been introduced in the main text, except $B_{ex}$, representing the exchange coupling between SLs.

By combining (\ref{eqn:C2}) and (\ref{eqn:C3}), a new expression can be obtained:
\begin{eqnarray}
\fl -\frac{\dot q}{\Delta}\left(Q_1s_1+Q_2s_2\right)+\alpha_1s_1\dot\psi_1+\alpha_2s_2\dot\psi_2=\nonumber\\
    =\left(Q_1p_1+Q_2p_2\right)\frac{J_{FiM}}{\Delta}-\nonumber\\
    -\frac{1}{2}\mu_0\left(N_y-N_x\right)\left[M_{s,1}^2\sin 2\psi_1+M_{s,2}^2\sin 2\psi_2+2M_{s,1}M_{s,2}\sin\left(\psi_1+\psi_2\right)\right]+\nonumber\\
    +\frac{\pi}{2}\left(\frac{Q_1D_1}{\Delta}\sin\psi_1+\frac{Q_2D_2}{\Delta}\sin\psi_2\right)\mathrm{.}\label{eqn:C4}
\end{eqnarray}

Now it is time to apply the approximations mentioned in the main text, that is, $Q=Q_1=-Q_2$ and $\psi_2+\pi\approx\psi_1=\psi$ ($Q=Q_1=Q_2$ and $\psi_2\approx\psi_1=\psi$) for the antiferromagnetic
(ferromagnetic) case. Only the antiferromagnetic case will be considered here (the ferromagnetic one can be straightforwardly derived). In this case, (\ref{eqn:C1}) and (\ref{eqn:C4}) can be rewritten as:
\begin{eqnarray}
\fl \frac{\dot q}{\Delta}\left(\alpha_1s_1+\alpha_2s_2\right)+Q\delta_s\dot\psi=\nonumber\\
    =-\left(\beta_1p_1+\beta_2p_2\right)\frac{J_{FiM}}{\Delta}+QMB_z-Q\frac{\pi}{2}\left(q_1+q_2\right)\cos\psi\frac{J_{HM}}{t_{FiM}}\mathrm{,}\label{eqn:C5}\\
\fl -Q\delta_s\frac{\dot q}{\Delta}+\left(\alpha_1s_1+\alpha_2s_2\right)\dot\psi=\nonumber\\
    =Q\left(p_1-p_2\right)\frac{J_{FiM}}{\Delta}-\frac{1}{2}\mu_0\left(N_y-N_x\right)\left(M_{s,1}-M_{s,2}\right)^2\sin 2\psi+\nonumber\\
    +Q\left(M_{s,1}-M_{s,2}\right)\frac{\pi}{2}\frac{D_1+D_2}{\left(M_{s,1}-M_{s,2}\right)\Delta}\sin\psi\mathrm{.}\label{eqn:C6}
\end{eqnarray}
Equations (\ref{eqn:C5}) and (\ref{eqn:01}) are directly comparable. By comparison of equation (\ref{eqn:C6}) and (\ref{eqn:02}), it follows the definition of the demagnetizing and iDMI fields as $B_m=\mu_0\left(N_x-N_y\right)M$ and $B_D=\frac{\pi}{2}\frac{D_1+D_2}{M\Delta}$, $M=M_{s,1}-M_{s_2}$ being the net saturation magnetization. These definitions are consistent with the equivalent expressions for pure ferromagnets.

\section{Determination of the DW width}\label{app:04}
The $\sigma_m$ value calculated in \ref{app:01} also determines the analytical calculation of the DW width. In this case, the whole functional must minimized with respect to $\Delta$. Under the approximations above in the antiferromagnetic coupling case, the derivative of $\sigma_m$ with respect to $\Delta$ yields:
\begin{equation}
\frac{\partial\sigma_m}{\partial\Delta}=\mu_0\left(M_{s,1}-M_{s,2}\right)^2\left(N_x\cos^2\psi+N_y\sin^2\psi-N_z\right)\mathrm{,}
\end{equation}
which again is consistent with the corresponding expression for pure ferromagnets. Hence, it can be obtained that:
\begin{equation}
\fl\Delta=\sqrt{\frac{A_1+A_2}{k_{u,1}+k_{u,2}-\frac{1}{2}\mu_0\left(M_{s,1}-M_{s,2}\right)^2\left[\left(N_z-N_x\right)+\left(N_x-N_y\right)\sin^2\psi\right]}}\mathrm{,}
\end{equation}
$A_i$ and $k_{u,i}$ being respectively the exchange and uniaxial out-of-plane anisotropy constants for each SL. A slightly more accurate version of this expression can be found in the literature.\cite{STejerina:20} Alternatively, $\Delta$ can be included among the parameters of the model.

\section*{References}
\providecommand{\noopsort}[1]{}\providecommand{\singleletter}[1]{#1}%

\end{document}